\documentclass[aps,prd,tightenlines,preprint,nofootinbib,a4paper]{revtex4}
\usepackage{epsfig}

\begin{document}

\title{Renormalization of the quark mass matrix}

\author{S. H. Chiu\footnote{schiu@mail.cgu.edu.tw}}
\affiliation{Physics Group, CGE, Chang Gung University, 
Taoyuan 33302, Taiwan}

\author{T. K. Kuo\footnote{tkkuo@purdue.edu}}
\affiliation{Department of Physics, Purdue University, West Lafayette, IN 47907, USA}

\begin{abstract}

Using a set of rephasing-invariant variables, it is shown that the renormalization
group equations for quark mixing parameters can be written in a form
that is compact, in addition to having simple properties under flavor permutation.
We also found approximate solutions to these equations if the quark masses
are hierarchical or nearly degenerate.

\end{abstract}


\maketitle

\pagenumbering{arabic}



\section{Introduction}

With the recent discovery of the Higgs boson, the last ``missing piece'' of the standard model
(SM) was finally found.  However, the long-standing mystery, that the Higgs couplings
(mass matrices) appear to be rather arbitrary, remains to be resolved.
A commonly held view posits that the SM is but an effective theory originating from
some other theory valid at high energies, and that more regularity can be found there.
To bridge these two energy regimes, one makes use of the renormalization group equations (RGEs).
Such RGEs for the mass matrices have been around for a long 
time (see, e.g., Refs.\cite{Cheng:1973nv,Ma:1979cw,Pendleton:1980as,
Hill:1980sq,Machacek:1983fi,Sasaki:1986jv,
Babu:1987im,Olechowski:1990bh,Barger:1992pk}).  
They are relatively simple when written in terms of the mass matrices themselves. 
However, these matrices contain a large number of unphysical degrees of freedom,
which must be stripped away to reveal the values of the physical variables, viz., 
the masses and the mixing matrices.
The procedure is by no means easy, and it is hard to correlate the variables in the
two energy regions.  For this reason a lot of efforts have gone into
recasting the RGEs into equations 
containing only physical variables \cite{Sasaki:1986jv,
Babu:1987im,Olechowski:1990bh,Barger:1992pk}.  With these equations       
the physical variables at different energies can be directly related.
Thus, for instance, one may test possible scenarios for mass patterns at high energies, 
using the RGE to see if they could evolve into the existing low-energy values.  
The challenge here comes from the complexity of the RGEs, which are lengthy, 
nonlinear, partial differential equations, so that the relations of variables at 
different energy scales are often obscure, and one can have only a partial view with
the use of various approximation schemes.  This difficulty, one would hope, can be mitigated
to some extent by a judicious choice of the physical variables.
Indeed, in this paper we propose to cast the one-loop quark RGEs in terms of a set of
rephasing-invariant variables 
introduced earlier \cite{Kuo:2005pf}. 
It is found that these RGEs
can be written in a compact form.  In addition, they exhibit manifest symmetries
which, as a consequence of the permutation properties of the chosen variables,
give these equations a very simple structure.  As it turns out,
this set of equations is still too complicated to be solved analytically.
However, under reasonable assumptions (hierarchy, degeneracy, etc.), approximate 
solutions are available.  These will be presented in this paper.
As more properties are found about these equations, one may hope that
they will help in the search for a viable high-energy theory.

\section{Rephasing-invariant parametrization}  

It is well known that physical observables are independent of rephasing
transformations on the mixing matrices of quantum-mechanical states.
Thus, instead of individual elements of the mixing matrix,
only rephasing-invariant combinations thereof are physical.
Whereas there is nothing wrong with using these elements in intermediate steps 
of a calculation, at the end of the day, they must form rephasing-invariant
combinations in physical quantities.
This situation is similar to that in gauge theory, where one often resorts
to a particular gauge choice for certain problems.
The final results, however, must be gauge invariant.  In this paper,
we propose to use, from the outset, parameters that are rephasing invariant.
As we will demonstrate in Sec. III, in terms of these, the quark RGEs become
quite simple in structure, making it easier to analyze the properties of 
their solutions.

We turn now to Ref.\cite{Kuo:2005pf}, 
where it was pointed out that six rephasing-invariant combinations can be constructed 
from elements of the Cabibbo-Kobayashi-Maskawa (CKM) matrix, $V$:
\begin{equation}\label{gamma}
\Gamma_{ijk}=V_{1i}V_{2j}V_{3k}=R_{ijk}-iJ,
\end{equation}
where $(i,j,k)$ is a cyclic permutation of $(1,2,3)$ and det$V=+1$ is imposed.
The common imaginary part is identified with the Jarlskog invariant \cite{Jarlskog:1985ht},
and the real parts are defined as
\begin{equation}\label{Rijk}
(R_{123},R_{231},R_{312};R_{132},R_{213},R_{321})=(x_{1},x_{2},x_{3};y_{1},y_{2},y_{3}).
\end{equation}
The $(x_{i},y_{j})$ parameters are bounded, $-1\leq (x_{i},y_{j}) \leq 1$,
with $x_{i} \geq y_{j}$ for any pair of $(i,j)$.
It is also found that the six parameters satisfy two conditions,
\begin{equation}\label{con1}
det V=(x_{1}+x_{2}+x_{3})-(y_{1}+y_{2}+y_{3})=1,
\end{equation}
\begin{equation}\label{con2}
(x_{1}x_{2}+x_{2}x_{3}+x_{3}x_{1})-(y_{1}y_{2}+y_{2}y_{3}+y_{3}y_{1})=0,
\end{equation}
leaving four independent parameters for the mixing matrix.
They are related to the Jarlskog invariant,
\begin{equation}
J^{2}=x_{1}x_{2}x_{3}-y_{1}y_{2}y_{3}.
\end{equation}
and the squared elements of $V$,
\begin{equation}
W=[|V_{\alpha i}|^{2}]=
\left(\begin{array}{ccc}
   x_{1}-y_{1} & x_{2}-y_{2} & x_{3}-y_{3} \\
   x_{3}-y_{2} & x_{1}-y_{3} & x_{2}-y_{1} \\
    x_{2}-y_{3} & x_{3}-y_{1} &x_{1}-y_{2} \\
    \end{array}
    \right)
    \end{equation}
The matrix of the cofactors of $W$, denoted as $w$ with $w^{T}W= (\mbox{det}W)I$, is given by
\begin{equation}
w=
\left(\begin{array}{ccc}
   x_{1}+y_{1} & x_{2}+y_{2} & x_{3}+y_{3} \\
   x_{3}+y_{2} & x_{1}+y_{3} & x_{2}+y_{1} \\
    x_{2}+y_{3} & x_{3}+y_{1} &x_{1}+y_{2} \\
    \end{array}
    \right)
    \end{equation}
The elements of $w$ are also bounded, $-1 \leq w_{\alpha i} \leq +1$, and
\begin{equation}
\sum_{i}w_{\alpha i}=\sum_{\alpha}w_{\alpha i}=\mbox{det} W,
\end{equation}
\begin{equation}
\mbox{det} W=\sum x_{i}^{2}-\sum y_{j}^{2}=\sum x_{i}+\sum y_{j}.
\end{equation}   
The relations between $(x_{i},y_{j})$ and the standard parametrization
can be found in Ref.\cite{Chiu:2015ega}.

There are some useful expressions for the rephasing-invariant combinations.
One first considers the product of four mixing elements \cite{Jarlskog:1985ht}
\begin{equation}\label{eq:piij}
\pi_{ij}^{\alpha \beta}=V_{\alpha i}V_{\beta j}V_{\alpha j}^{*}V_{\beta i}^{*},
\end{equation}
which can be reduced to 
\begin{eqnarray}
\pi_{ij}^{\alpha \beta} & = & |V_{\alpha i}|^{2}|V_{\beta j}|^{2}-
 \sum_{\gamma k}\epsilon _{\alpha \beta \gamma}\epsilon_{ijk}V_{\alpha i}V_{\beta j}V_{\gamma k} \nonumber \\
   & = & |V_{\alpha j}|^{2}|V_{\beta i}|^{2}+
 \sum_{\gamma k}\epsilon _{\alpha \beta \gamma}\epsilon_{ijk}V_{\alpha j}^{*}V_{\beta i}^{*}V_{\gamma k}^{*},
 \end{eqnarray}
In addition, for $\alpha \neq \beta \neq \gamma$ and
$i\neq j \neq k$, we define
\begin{equation}\label{piab}
\pi^{\alpha \beta}_{ij}\equiv \pi_{\gamma k}=\Lambda_{\gamma k}+iJ.
\end{equation}
Since $Re(\pi^{\alpha \beta}_{ij})$ takes the forms,
\begin{equation}
Re(\pi^{\alpha \beta}_{ij})=|V_{\alpha i}|^{2}|V_{\beta j}|^{2}-x_{a}=
|V_{\beta i}|^{2}|V_{\alpha j}|^{2}+y_{b},
\end{equation}
we have
\begin{equation}\label{lambda}
\Lambda_{\gamma k}=
\frac{1}{2}(|V_{\alpha i}|^{2}|V_{\beta j}|^{2}+|V_{\alpha j}|^{2}|V_{\beta i}|^{2}-|V_{\gamma k}|^{2}).
\end{equation}
In terms of the $(x,y)$ variables,
\begin{equation}\label{lambdaxy}
\Lambda_{\gamma k}=x_{a}y_{j}+x_{b}x_{c}-y_{j}(y_{k}+y_{l}),
\end{equation}
where $(x_{a},y_{j})$ comes from $|V_{\gamma k}|^{2}=x_{a}-y_{j}$, 
and $a\neq b\neq c$, $j\neq k \neq l$.


\begin{table}\label{tZ}
 \begin{center}
 \begin{tabular}{ccc}
   \\     \hline \hline  
\vspace{0.15in}   
  $Z_{1}=\left(\begin{array}{ccc}
   \Lambda_{11} & 0 & 0 \\
    0 & \Lambda_{22} & 0 \\
    0 &0 &\Lambda_{33} \\
    \end{array}
    \right)$, 
 &  $Z_{2}=\left(\begin{array}{ccc}
   0 & \Lambda_{12} & 0 \\
    0 & 0 & \Lambda_{23} \\
    \Lambda_{31}& 0 &0 \\
    \end{array}
    \right)$, 
 
    &  $Z_{3}=\left(\begin{array}{ccc}
   0 & 0 & \Lambda_{13} \\
   \Lambda_{21} & 0 & 0 \\
    0 &\Lambda_{32} & 0 \\
    \end{array}
    \right)$ 
 
 \\ \vspace{.15in} 
   $Z'_{1}=\left(\begin{array}{ccc}
   \Lambda_{11} & 0 & 0 \\
   0 & 0 & \Lambda_{23} \\
    0 &\Lambda_{32} &0 \\
    \end{array}
    \right)$,    
  &  $Z'_{2}=\left(\begin{array}{ccc}
   0 & \Lambda_{12} & 0 \\
    \Lambda_{21} & 0 & 0 \\
   0 & 0 &\Lambda_{33} \\
    \end{array}
    \right)$,  
   
  &  $Z'_{3}=\left(\begin{array}{ccc}
   0 & 0 & \Lambda_{13} \\
    0 & \Lambda_{22} & 0 \\
    \Lambda_{31} &0 &0 \\
    \end{array}
    \right)$     
   \\  \hline 
   \end{tabular}
  \begin{tabular}{ccc}
\vspace{0.25in}

  $[Z_{0}]$ & = & $\left(\begin{array}{ccc}
   (1-|V_{11}|^{2})\Lambda_{11} & (1-|V_{12}|^{2})\Lambda_{12} & (1-|V_{13}|^{2})\Lambda_{13} \\
    (1-|V_{21}|^{2})\Lambda_{21} & (1-|V_{22}|^{2})\Lambda_{22} & (1-|V_{23}|^{2})\Lambda_{23} \\
    (1-|V_{31}|^{2})\Lambda_{31} & (1-|V_{32}|^{2})\Lambda_{32} & (1-|V_{33}|^{2})\Lambda_{33} \\
    \end{array}
    \right)$  \\

   \hline \hline
   
    \end{tabular}
 \caption{The explicit expressions of the matrices $[Z_{i}]$, $[Z'_{i}]$,
 and $[Z_{0}]$. Here $\Lambda _{\gamma k}$ 
is defined in Eq.~(\ref{lambda}).} 
   \end{center}
 \end{table}  

\section{RGEs for quarks} 

The one-loop RGEs for the quark mass matrices
have been developed and widely studied \cite{Machacek:1983fi,Sasaki:1986jv,
Babu:1987im}.  In terms of  
the mass-squared matrices for the $u$-type quarks, $M_{u}=Y_{u}Y^{\dag}_{u}$, 
and that for the $d$-type quarks,
$M_{d}=Y_{d}Y^{\dag}_{u}$, 
where $Y$ is the Yukawa coupling matrices of the Higgs boson to the quarks,
the RGEs take a simple form:
\begin{equation}\label{eq:Mu}
\mathcal{D} M_{u}=a_{u}M_{u}+bM^{2}_{u}+c\{M_{u},M_{d}\},
\end{equation}  
\begin{equation}\label{eq:Md}
\mathcal{D}M_{d}=a_{d}M_{d}+bM^{2}_{d}+c\{M_{u},M_{d}\}.
\end{equation}
Here, $\mathcal{D} \equiv (16\pi^{2})\frac{d}{dt}$ and $t=\ln(\mu/M_{W})$,
where $\mu$ is the energy scale and $M_{W}$ is the $W$ boson mass.
The model dependence of the RGEs is implanted in $a_{u}$, $a_{d}$, $b$, and $c$.


\begin{table}\label{tS}
 \begin{center}
 \begin{tabular}{ccc}
   \\     \hline \hline  
\vspace{0.15in}   
  $S_{11}=\left(\begin{array}{ccc}
   0 & 0 & 0 \\
    0 & \Lambda_{22} & -\Lambda_{23} \\
    0 &-\Lambda_{32} &\Lambda_{33} \\
    \end{array}
    \right)$, 
 &  $S_{12}=\left(\begin{array}{ccc}
   0 & 0 & 0 \\
    -\Lambda_{21} & 0 & \Lambda_{23} \\
    \Lambda_{31} & 0 &-\Lambda_{33} \\
    \end{array}
    \right)$, 
 
    &  $S_{13}=\left(\begin{array}{ccc}
   0 & 0 & 0 \\
   \Lambda_{21} & -\Lambda_{22} & 0 \\
    -\Lambda_{31} &\Lambda_{32} & 0 \\
    \end{array}
    \right)$ 
 
 \\ \vspace{.15in} 
   $S_{21}=\left(\begin{array}{ccc}
   0 & -\Lambda_{12} & \Lambda_{13} \\
   0 & 0 & 0 \\
    0 &\Lambda_{32} &-\Lambda_{33} \\
    \end{array}
    \right)$,    
  &  $S_{22}=\left(\begin{array}{ccc}
   \Lambda_{11} & 0 & -\Lambda_{13} \\
    0 & 0 & 0 \\
    -\Lambda_{31} & 0 &\Lambda_{33} \\
    \end{array}
    \right)$,  
   
  &  $S_{23}=\left(\begin{array}{ccc}
   -\Lambda_{11} & \Lambda_{12} & 0 \\
    0 & 0 & 0 \\
    \Lambda_{31} &-\Lambda_{32} &0 \\
    \end{array}
    \right)$ 
 
 \\ \vspace{.15in} 
   $S_{31}=\left(\begin{array}{ccc}
   0 & \Lambda_{12} & -\Lambda_{13} \\
    0 & -\Lambda_{22} & \Lambda_{23} \\
    0 & 0 & 0 \\
    \end{array}
    \right)$,   
  &  $S_{32}=\left(\begin{array}{ccc}
   -\Lambda_{11} & 0 & \Lambda_{13} \\
    \Lambda_{21} & 0 & -\Lambda_{23} \\
    0 & 0 & 0 \\
    \end{array}
    \right)$, 
   
  &  $S_{33}=\left(\begin{array}{ccc}
   \Lambda_{11} & -\Lambda_{12} & 0 \\
   -\Lambda_{21} & \Lambda_{22} & 0 \\
    0 & 0 & 0 \\
    \end{array}
    \right)$    
   \\  \hline \hline
   
    \end{tabular}
 \caption{The explicit expressions of the matrix $[S_{ij}]$.} 
   \end{center}
 \end{table}


Although the RGEs are simple in their matrix forms,
one must extract the physical variables (masses and mixing parameters)
from these matrices.  This is complicated because they contain a large number of
unphysical degrees of freedom and it is not easy to infer the evolution of
the physical variables from that of the mass matrices. 
For this reason it is useful to deduce from Eqs.~(\ref{eq:Mu}-\ref{eq:Md})
the RGEs in terms of the physical variables, which can then yield direct information 
on the evolution of these variables.  This procedure results in the following
equations for the masses and CKM elements $V_{ij}$: 

\begin{equation}
\mathcal{D} \ln(f^{2}_{i})=a_{u}+bf^{2}_{i}+2c\sum_{j}h^{2}_{j}|V_{ij}|^{2},
\end{equation}
\begin{equation}
\mathcal{D} \ln(h^{2}_{i})=a_{d}+bh^{2}_{i}+2c\sum_{j}f^{2}_{j}|V_{ij}|^{2},
\end{equation}
\begin{equation}\label{eq:vij}
\mathcal{D}V_{ij}=c[\sum_{l,k\neq i}F_{ik}h^{2}_{l}V_{il}V^{*}_{kl}V_{kj}+
\sum_{m,k\neq j} H_{jk}f^{2}_{m}V^{*}_{mk}V_{mj}V_{ik}],
\end{equation}
where $f^{2}_{i}$ and $h^{2}_{i}$ are the eigenvalues of $M_{u}$ and $M_{d}$, respectively, and
\begin{equation}
F_{ik}=\frac{f^{2}_{i}+f^{2}_{k}}{f^{2}_{i}-f^{2}_{k}}, \: \: \:
H_{jk}=\frac{h^{2}_{j}+h^{2}_{k}}{h^{2}_{j}-h^{2}_{k}},
\end{equation}

It should be emphasized that Eq.~(\ref{eq:vij}), as it stands, is not rephasing-invariant.
The physical part thereof is obtained by using it only on
rephasing invariant combinations of $V_{ij}$, such as $|V_{ij}|^{2}$
or the $(x,y)$ variables defined in Eq.~(\ref{Rijk}). 
In Ref.\cite{Chiu:2008ye}, we obtained the evolution              
equations of $x_{i}$ and $y_{j}$ in the form 
\begin{eqnarray}\label{xi}
-\mathcal{D}x_{i}/c & = & [\Delta f^{2}_{23}, \Delta f^{2}_{31}, \Delta f^{2}_{12}][A_{i}]
                            [H_{23},H_{31},H_{12}]^{T} \nonumber \\
                    & + & [\Delta h^{2}_{23}, \Delta h^{2}_{31}, \Delta h^{2}_{12}][B_{i}]
                            [F_{23},F_{31},F_{12}]^{T},
\end{eqnarray}
\begin{eqnarray}\label{yi}
-\mathcal{D}y_{i}/c & = & [\Delta f^{2}_{23}, \Delta f^{2}_{31}, \Delta f^{2}_{12}][A'_{i}]
                            [H_{23},H_{31},H_{12}]^{T} \nonumber \\
                    & + & [\Delta h^{2}_{23}, \Delta h^{2}_{31}, \Delta h^{2}_{12}][B'_{i}]
                            [F_{23},F_{31},F_{12}]^{T},
\end{eqnarray}
where $\Delta f^{2}_{ij}=f^{2}_{i}-f^{2}_{j}$ and $\Delta h^{2}_{ij} =h^{2}_{i}-h^{2}_{j}$.
In terms of $(x_{i},y_{j})$, the
explicit forms of the matrices $[A_{i}]$, $[A'_{i}]$, $[B_{i}]$, and $[B'_{i}]$ are given in 
Table II of Ref.\cite{Chiu:2008ye}.  Since $\sum \Delta f^{2}_{ij}=\sum \Delta h^{2}_{ij}=0$, 
to the matrices
$[A_{i}]$, $[B_{i}]$, $[A'_{i}]$, and $[B'_{i}]$, we can add arbitrary matrices of the form
\begin{eqnarray}
\left(\begin{array}{ccc}
   \delta_{1} & \delta_{2} & \delta_{3}  \nonumber \\
   \delta_{1} &\delta_{2} & \delta_{3} \nonumber \\
    \delta_{1} & \delta_{2} & \delta_{3} \nonumber \\
    \end{array} \nonumber
    \right).   \nonumber
\end{eqnarray}
Thus, for instance, from Table II in Ref.\cite{Chiu:2008ye}
\begin{eqnarray}
[A_{1}] &=& \left(\begin{array}{ccc}
   2x_{1}y_{1} & x_{1}x_{2}+y_{2}y_{3} & x_{1}x_{3}+y_{2}y_{3} \nonumber \\
   x_{1}x_{3}+y_{1}y_{2} &  2x_{1}y_{3} & x_{1}x_{2}+y_{1}y_{2} \nonumber \\
    x_{1}x_{2}+y_{1}y_{3} & x_{1}x_{3}+y_{1}y_{3} & 2x_{1}y_{2} \nonumber \\
    \end{array} \nonumber
    \right)   \nonumber \\
& = & 2[Z_{1}]-[Z_{0}]+(J^{2}+3\sum x_{i}x_{j}-x_{2}x_{3})
\left(\begin{array}{ccc}
   1 & 1 & 1   \\
   1 & 1 & 1  \\
    1 & 1 & 1  \\
    \end{array}
    \right)
-\left(\begin{array}{ccc}
   y_{2}y_{3} & y_{1}y_{2} & y_{1}y_{3}   \\
   y_{2}y_{3} & y_{1}y_{2} & y_{1}y_{3}  \\
   y_{2}y_{3} & y_{1}y_{2} & y_{1}y_{3} \\
    \end{array}
    \right),
    \end{eqnarray}
where we have used the relations $W_{KL}\Lambda_{KL}=J^{2}+x_{a}y_{j}$,
$W_{KL}=x_{a}-y_{j}$.
It follows that
\begin{equation}
[\Delta f^{2}_{23},\Delta f^{2}_{31},\Delta f^{2}_{12}][A_{1}]=
[\Delta f^{2}_{23},\Delta f^{2}_{31},\Delta f^{2}_{12}](2[Z_{1}]-[Z_{0}]).
    \end{equation}
Similarly, all the $[A]$ and $[B]$ matrices can be so transformed and we may recast
Eqs.~(\ref{xi}-\ref{yi}) in a more suggestive form,
\begin{eqnarray}\label{Dxi}
-\mathcal{D}x_{i}/c &=&[\Delta f^{2}_{23}, \Delta f^{2}_{31}, \Delta f^{2}_{12}]
(2[Z_{i}]-[Z_{0}])[H_{23},H_{31},H_{12}]^{T} \nonumber \\
&+&[\Delta h^{2}_{23}, \Delta h^{2}_{31}, \Delta h^{2}_{12}]
(2[Z_{i}]-[Z_{0}])^{T}[F_{23},F_{31},F_{12}]^{T},
\end{eqnarray}
\begin{eqnarray}\label{Dyi}
-\mathcal{D}y_{i}/c &=&[\Delta f^{2}_{23}, \Delta f^{2}_{31}, \Delta f^{2}_{12}]
(2[Z'_{i}]-[Z_{0}])[H_{23},H_{31},H_{12}]^{T} \nonumber \\
&+&[\Delta h^{2}_{23}, \Delta h^{2}_{31}, \Delta h^{2}_{12}]
(2[Z'_{i}]-[Z_{0}])^{T}[F_{23},F_{31},F_{12}]^{T}.
\end{eqnarray}
The matrices $[Z_{i}]$, $[Z'_{i}]$, and $[Z_{0}]$ are listed in Table I.   
It is noteworthy that the matrix structures of $[Z_{i}]$ and $[Z'_{i}]$
mirror those of $x_{i}$ and $y_{i}$, when written as products of $V_{ij}$,
e.g., $x_{1}=Re(V_{11}V_{22}V_{33})$.  It is also satisfying to
establish $[B_{i}]=[A_{i}]^{T}$ and $[B'_{i}]=[A'_{i}]^{T}$,
which is a consequence of the conjugate roles played by the $u$-type
and $d$-type quarks.  The RGEs of $W_{ij}(|V_{ij}|^{2})$ and $J^{2}$
can be obtained:

\begin{eqnarray}\label{DWij}
-\frac{1}{2c}\mathcal{D}W_{ij} & =  & [\Delta f^{2}_{23}, \Delta f^{2}_{31}, \Delta f^{2}_{12}][S_{ij}]
                            [H_{23},H_{31},H_{12}]^{T} \nonumber \\
                    & + & [\Delta h^{2}_{23}, \Delta h^{2}_{31}, \Delta h^{2}_{12}][S_{ij}]^{T}
                            [F_{23},F_{31},F_{12}]^{T},
\end{eqnarray}
\begin{eqnarray}\label{lnJ}
-\frac{1}{2c}\mathcal{D}\ln J^{2}/c & = & [\Delta f^{2}_{23}, \Delta f^{2}_{31}, \Delta f^{2}_{12}][w]
                            [H_{23},H_{31},H_{12}]^{T} \nonumber \\
                    & + & [\Delta h^{2}_{23}, \Delta h^{2}_{31}, \Delta h^{2}_{12}][w]^{T}
                            [F_{23},F_{31},F_{12}]^{T}.
\end{eqnarray}
Although $[S_{ij}]$ can be directly written down from $[Z_{i}]$
and $[Z'_{i}]$, we list them explicitly in Table II, since it will be   
used for the analyses of $\mathcal{D} W_{ij}$ in the next section.

The simple and compact form of Eqs.~(\ref{Dxi}-\ref{lnJ}) can be
contrasted with the RGEs written in terms of the standard 
parametrization (see, e.g., Ref.\cite{Balzereit:1998id}), 
for which it is hard to find any regularity in the structure.
It is seen that these equations clearly exhibit symmetries under permutation
of the indices, owing to the same properties inherent in the definition of
the $(x,y)$ variables.  
The situation here can be compared to a familiar one in electricity and magnetism.  
While the wave equations
take a simple form for the (gauge-invariant) $\vec{E}$ and $\vec{B}$ fields,
depending on the choice of gauge, the corresponding equations
for the potential $A_{\mu}$ can be very complicated.
Another salient feature of them is the prominent 
role played by the rephasing invariants $\Lambda_{\gamma k}$,
which are the same Jarlskog invariants that appear in formulas of the
neutrino oscillation probabilities, $P(\nu_{\alpha} \rightarrow \nu_{\beta})$.
Without them the RGEs would look rather cumbersome,
as written in Ref.\cite{Chiu:2008ye}.  In addition, they facilitate the calculation of 
approximate solutions of the RGEs, as we will see in the next section.  
Last, from Eqs.~(\ref{Dxi}), ~(\ref{Dyi}), and Table I,  
it can be verified that $\sum \mathcal{D}(x_{i})-\sum \mathcal{D}(y_{j})=0$
and $\sum \mathcal{D}(x_{i}x_{j})-\sum \mathcal{D}(y_{i}y_{j})=0$,
as one expects from the constraint equations [Eqs.~(\ref{con1}) and ~(\ref{con2})].

Notice that the evolution equations of $\Lambda_{\gamma k}$ can also be cast in 
compact forms similar to that of $W_{ij}$ and $J^{2}$:
\begin{eqnarray}\label{Lambda}
-\frac{1}{2c}\mathcal{D}\Lambda_{\gamma k} & =  & [\Delta f^{2}_{23}, \Delta f^{2}_{31}, \Delta f^{2}_{12}][Y_{\gamma k}]
                            [H_{23},H_{31},H_{12}]^{T} \nonumber \\
                    & + & [\Delta h^{2}_{23}, \Delta h^{2}_{31}, \Delta h^{2}_{12}][Y_{\gamma k}]^{T}
                            [F_{23},F_{31},F_{12}]^{T}.
\end{eqnarray}
Here the matrix $[Y_{\gamma k}]$ takes the form
\begin{equation}
[Y_{\gamma k}]=
\left(\begin{array}{ccc}
   c_{11}\Lambda_{11} & c_{12}\Lambda_{12} & c_{13}\Lambda_{13} \\
   c_{21}\Lambda_{21} & c_{22}\Lambda_{22} & c_{23}\Lambda_{23} \\
    c_{31}\Lambda_{31} & c_{32}\Lambda_{32} & c_{33}\Lambda_{33} \\
    \end{array}
    \right),
\end{equation}
where the coefficients $c_{ij}$ are functions of $|V_{ij}|^{2}$.  As an example,
\begin{equation}
[Y_{11}]=
\left(\begin{array}{ccc}
   (|V_{23}|^{2}+|V_{32}|^{2}-|V_{22}|^{2}-|V_{33}|^{2})\Lambda_{11} 
   & (|V_{22}^{2}-|V_{32}|^{2})\Lambda_{12} 
   & (|V_{33}|^{2}-|V_{23}|^{2})\Lambda_{13} \\
   (|V_{22}|^{2}-|V_{23}|^{2})\Lambda_{21} & (1-|V_{22}|^{2})\Lambda_{22} & (-1+|V_{23}|^{2})\Lambda_{23} \\
    (|V_{33}|^{2}-|V_{32}|^{2})\Lambda_{31} & (-1+|V_{32}|^{2})\Lambda_{32} & (1-|V_{33}|^{2})\Lambda_{33} \\
    \end{array}
    \right).
\end{equation}
It is seen that 
\begin{equation}\label{sumcij}
\sum_{i}c_{Ii}=\sum_{I}c_{Ii}=0,  
\end{equation}
and the $2\times 2$ submatrix
(indices 2 and 3) has a simple structure, $c_{\gamma k}=\pm 1 \pm |V_{\gamma k}|^{2}$,
$(\gamma k)=(2,3)$.  With the condition, Eq.~(\ref{sumcij}), one can construct the
$3\times 3$ matrix from the known $2\times 2$ matrix.
Finally, the evolution equations for the combinations of $\Lambda_{\gamma k}$,
such as $\mathcal{D}(\sum_{\gamma}\Lambda_{\gamma k})$, $\mathcal{D}(\sum_{k}\Lambda_{\gamma k})$,
and $\mathcal{D}(\sum_{\gamma,k}\Lambda_{\gamma k})$, can also be cast in 
similar forms, in which $c_{ij}$ are functions of the elements of $W_{ij}$ and $w_{ij}$.
We will not show the details here.


\section{Analysis of the RGE}

Although the solutions to the quark RGEs are not available,
it turns out that, under certain reasonable assumptions, one can find
approximate solutions for them.  
Before embarking on this analysis, it should be noticed that, with 
the observed values in the mass matrices, the parameter $c/16\pi^{2}$ and all $\Lambda_{ij}$'s are small.  
This means that renormalization effects are generally small if one starts from low
energy using the SM and the known values of the  physical variables.  However, it is
interesting to entertain the possibility that, at some point, a new theory can intervene
with a fast-paced renormalization evolution.  It is then relevant to consider RGE
evolution from high to low $t$ values, with other assumed parameters at high energies.  
To do this we consider various scenarios
of the mass parameters:
A) $f_{3}^{2} \gg f_{2}^{2} \gg f_{1}^{2}$ and $h_{3}^{2} \gg h_{2}^{2} \gg h_{1}^{2}$;
B) $f_{3}^{2} \gg f_{2}^{2} \approx f_{1}^{2}$ 
and $h_{3}^{2} \gg h_{2}^{2} \approx h_{1}^{2}$;  
C) $f_{3}^{2} \gg f_{2}^{2} \gg f_{1}^{2}$ and $h_{3}^{2} \gg h_{2}^{2} \approx h_{1}^{2}$.
While case A) corresponds to the mass patterns at low energy, the other choices
are possibilities which may prevail at some high energy scale.
These considerations are useful for model building, so that one can bridge the mixing
patterns between the high and low energy scales.  We will now present the detailed
results for case A), but leave the discussion of the other cases to the Appendix.

For the hierarchical case in A),
one may simplify the matrices so that $[F_{23},F_{31},F_{12}]\simeq [-1,1,-1]$ and
$[H_{23},H_{31},H_{12}]\simeq [-1,1,-1]$.  In addition,
\begin{equation}
[\Delta f^{2}_{23},\Delta f^{2}_{31},\Delta f^{2}_{12}]\simeq f^{2}_{3}[-1,1,0],
\end{equation}
\begin{equation}
[\Delta h^{2}_{23},\Delta h^{2}_{31},\Delta h^{2}_{12}]\simeq h^{2}_{3}[-1,1,0].
\end{equation}
The approximations lead to
\begin{equation}\label{HH}
-\frac{1}{2c}\mathcal{D}W_{ij} \simeq (f^{2}_{3}[\sum_{p,q}(-1)^{p+q}S^{pq}_{ij}]
+h^{2}_{3}[\sum_{p,q}(-1)^{p+q}S^{pq}_{ij}]^{T}),
\end{equation}
where $S^{pq}_{ij}$ is the $(p,q)$ element of $S_{ij}$
with $p=1,2$ and $q=1,2,3$.
We show the explicit expressions of $\mathcal{D} W_{ij}$ in the Appendix.

Note that out of the nine equations, six of them can be cast in the following forms:
\begin{equation}
\frac{1}{2c}\mathcal{D}\ln W_{11}=f^{2}_{3}W_{31}+h^{2}_{3}W_{13},
\end{equation}
\begin{equation}
\frac{1}{2c} \mathcal{D}\ln W_{13}=-f^{2}_{3}W_{33}-h_{3}^{2}(1-W_{13}),
\end{equation}
\begin{equation}
\frac{1}{2c}\mathcal{D}\ln W_{23}=-f^{2}_{3}W_{33}-h^{2}_{3}(W_{33}-W_{13}),
\end{equation}
\begin{equation}
\frac{1}{2c}\mathcal{D}\ln W_{31}=-f^{2}_{3}(1-W_{31})-h^{2}_{3}W_{33},
\end{equation}
\begin{equation}
\frac{1}{2c}\mathcal{D}\ln W_{32}=-f^{2}_{3}(W_{33}-W_{31})-h^{2}_{3}W_{33},
\end{equation}
\begin{equation}
\frac{1}{2c}\mathcal{D}\ln W_{33}=f^{2}_{3}(1-W_{33})+h^{2}_{3}(1-W_{33}).
\end{equation}
A RGE invariant can then be derived directly,
\begin{equation}
\mathcal{D}\ln (\frac{W_{13}W_{31}W_{33}}{W_{23}W_{32}})=0.
\end{equation}

Since from the theoretical point of view there is no preferred scenario concerning the relative
magnitudes of $f^{2}_{i}$ and $h^{2}_{i}$ at high energies,
it would be interesting to further pursue possible invariants
under the following assumptions about the couplings.
(i) If $f^{2}_{3} \gg h^{2}_{3}$, we obtain three more approximate invariants:
\begin{equation}
\mathcal{D}\ln (\frac{W_{13}}{W_{23}})\simeq 0,
\end{equation}
\begin{equation}
\mathcal{D}\ln (\frac{W_{11}W_{13}}{W_{32}})\simeq 0,
\end{equation}
\begin{equation}
\mathcal{D}\ln (\frac{W_{31}W_{33}}{W_{32}})\simeq 0,
\end{equation}
(ii) If on the other hand, $f^{2}_{3} \ll h^{2}_{3}$, we have
\begin{equation}
\mathcal{D}\ln (\frac{W_{31}}{W_{32}})\simeq 0.
\end{equation}
\begin{equation}
\mathcal{D}\ln (\frac{W_{13}W_{33}}{W_{23}}) \simeq 0,
\end{equation}
\begin{equation}
\mathcal{D}\ln (\frac{W_{11}W_{31}}{W_{23}}) \simeq 0.
\end{equation}

\begin{figure}[ttt]
\caption{The approximate solutions (dashed) are compared with the full, numerical solutions (solid)
for the hierarchical scenario with $f^{2}_{3}=h^{2}_{3}=4$, 
where $f^{2}_{3} \gg f^{2}_{2} \gg f^{2}_{1}$ and $h^{2}_{3} \gg h^{2}_{2} \gg h^{2}_{1}$.
Here $(b,c)=(3,-3/2)$ under the standard model. 
The initial values of $(x,y)$ at $t=30$ are taken to be $x_{1}=(1/6)+\varepsilon$, 
$x_{2}=(1/6)-\varepsilon$, $y_{1}=-(1/6)+\varepsilon$, and $-(1/6)-\varepsilon$,
where $\varepsilon =0.01$} 
\centerline{\epsfig{file=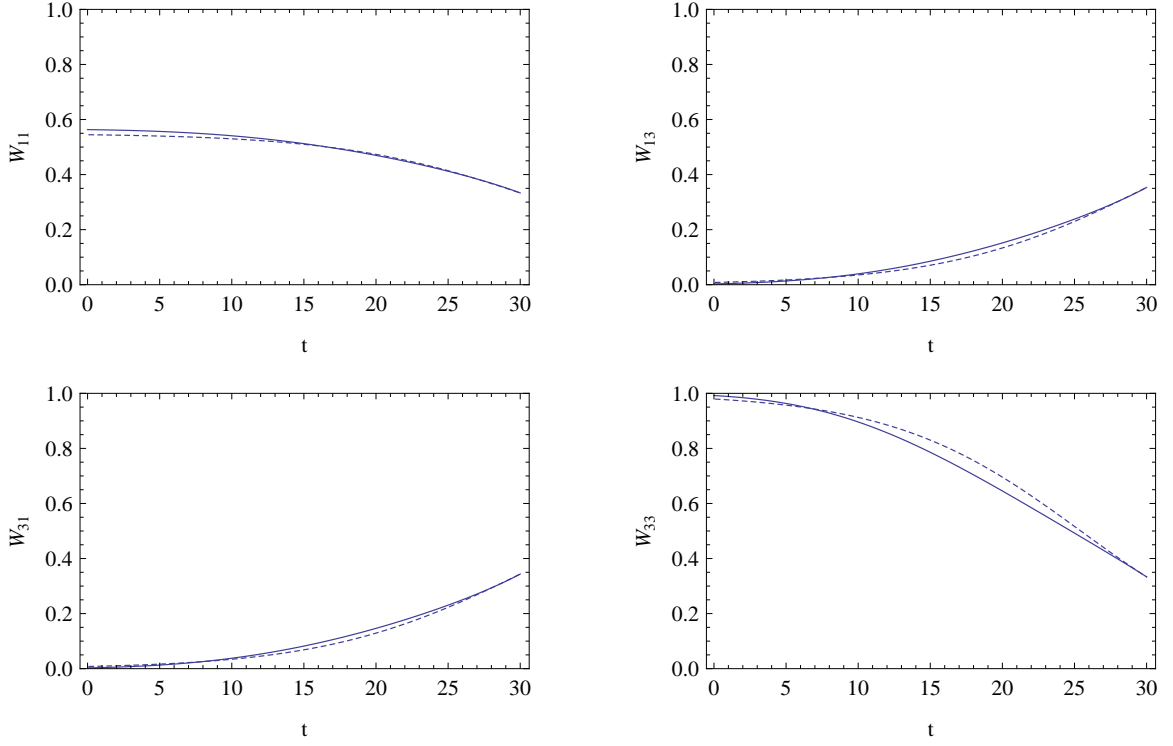,width= 16cm}}
\end{figure}

Despite the complexity of its original forms, 
the RGEs of $W_{ij}$ can be solved approximately. With $c'=16\pi^{2}/[2c(f^{2}_{3}+h^{2}_{3})]$
and $a_{ij}$ the initial value of $W_{ij}$,
Eq.~(\ref{DW33}) yields
\begin{equation}\label{W33}
W_{33}\simeq \frac{1}{(a^{-1}_{33}-1)e^{-(t-t_{0})/c'}+1}.
\end{equation}
With the solution of $W_{33}$, one may in principle solve for $W_{13}$, $W_{33}$, and
$W_{11}$.  However, we will not show the long expressions here, but instead further
assume the following scenarios of the couplings to obtain simple, 
approximate solutions for the rest of the $W_{ij}$.
Note that $f^{2}_{3}$ and $h^{2}_{3}$ are treated as constants here,
i.e., the approximate solutions are only valid for a range of $t$ values in which
the variations of $f^{2}_{3}$ and $h^{2}_{3}$ are negligible.
\begin{itemize}
\item If $f^{2}_{3} \gg h^{2}_{3}$, it leads to
   \begin{equation}
   W_{13}\simeq \frac{a_{13}}{(1-a_{33})+a_{33}e^{(t-t_{0})/c_{f}}},
   \end{equation}
   \begin{equation}
   W_{31}\simeq \frac{a_{31}}{a_{31}+(1-a_{31})e^{(t-t_{0})/c_{f}}},
   \end{equation}
   \begin{equation}
   W_{11}\simeq \frac{a_{11}}{(1-a_{31})+a_{31}e^{-(t-t_{0})/c_{f}}},
   \end{equation}
where $c_{f}=16\pi^{2}/(2cf^{2}_{3})\approx c'$.
\item If $f^{2}_{3} \ll h^{2}_{3}$, 
   \begin{equation}
   W_{13}\simeq \frac{a_{13}}{a_{13}+(1-a_{13})e^{(t-t_{0})/c_{h}}},
   \end{equation}
   \begin{equation}
   W_{31}\simeq \frac{a_{31}}{(1-a_{33})+a_{33}e^{(t-t_{0})/c_{h}}},
   \end{equation}
   \begin{equation}
   W_{11}\simeq \frac{a_{11}}{(1-a_{13})+a_{13}e^{-(t-t_{0})/c_{h}}},
   \end{equation}
where $c_{h}=16\pi^{2}/(2ch^{2}_{3})\approx c'$.
\item If $f^{2}_{3} \approx h^{2}_{3}$, then $c_{f}\approx c_{h} \approx 2c'$, and
   \begin{equation}
   W_{13}\simeq \frac{a_{13}(1-a_{33})}{\frac{a_{13}K}{1-L}+(1-a_{13}-a_{33})K},
   \end{equation}
   \begin{equation}
   W_{31}\simeq \frac{a_{31}(1-a_{33})}{\frac{a_{31}K}{1-L}+(1-a_{31}-a_{33})K},
   \end{equation}
   \begin{eqnarray}
   W_{11}&\simeq& a_{11}(1-a_{33})
   [-1+2a_{13}+a_{33}-a^{2}_{13}L]^{-1/2}[-1+2a_{31}+a_{33}-a^{2}_{31}L]^{-1/2} \nonumber \\
   & \cdot& [\frac{a_{13}(K+1)-(1-a_{33})}
   {a_{13}(K-1)+(1-a_{33})}]^{1/2} 
   \cdot  [\frac{a_{31}(K+1)-(1-a_{33})}
   {a_{31}(K-1)+(1-a_{33})}]^{1/2},
   \end{eqnarray}
where $L=1-\exp[-(t-t_{0})/c']$ and $K=\sqrt{1-L+a_{33}L}$. 

\end{itemize}

For the purpose of illustration, we show a numerical example in Fig. 1,
in which the approximate solutions for $W_{11}$, $W_{13}$, $W_{31}$, and $W_{33}$
are compared with the full numerical solutions.  
It is seen that although $f^{2}_{3}$ and $h^{2}_{3}$ are treated as constants in the 
approximation, the resultant solutions agree well with the full numerical
solutions in which $f^{2}_{3}$ and $h^{2}_{3}$ vary by a factor of $4$.
Note that due to a lack of details at the high energy regimes,
the chosen input at high-energy in this example only leads to 
$W_{11}\approx 3/5$ at low energy.


\section{conclusion}

One of the cornerstones of quantum field theories is the RGE of coupling
``constants'' which describe the change of couplings as functions of energy scales.
When applied to gauge couplings, they led to the well-established phenomenon of
asymptotic freedom, in addition to the concept of unification, which is a most
interesting conjecture for high-energy theories.  Given the plethora of
masses and mixing parameters, one would hope that RGEs can introduce some
regularity, or at least certain insights, into this set of seemingly random
observables.  However, so far this goal remains largely unfulfilled.
One obvious obstacle comes from the complexity of the RGEs, when written in
terms of the variables of the standard parametrization.
In this paper we obtained evolution equations for a set of rephasing-invariant
mixing parameters. They exhibit compact and simple structures,
with manifest permutation symmetry.  Although a full analysis of 
these equations is still lacking, they are simple enough for one to find
approximate solutions under a number of reasonable assumptions for
possible mass parameters.  They should be helpful in assessing the
viability of proposed theories at high energies.  Hopefully,
as we learn more about these equations, we can have a clear
picture of the relations of Higgs couplings between low and
high energies.

\acknowledgments                 
S.H.C. is supported by the Ministry of Science and Technology of Taiwan, 
Grant No.: MOST 104-2112-M-182-004.

\appendix

\section{}


Following the discussions in Sec. IV, in this appendix we collect the explicit RGEs
under various assumptions about the quark masses, whether hierarchical or nearly degenerate,
when appropriate, we also present approximate solutions for the individual cases.

\subsection{Case A): $f_{3}^{2} \gg f_{2}^{2} \gg f_{1}^{2}$ and 
$h_{3}^{2} \gg h_{2}^{2} \gg h_{1}^{2}$}

In this case, the explicit expressions of $\mathcal{D} W_{ij}$ following Eq.~(\ref{HH}) 
are given by
\begin{equation}
\frac{1}{2c}\mathcal{D}W_{11} \simeq f^{2}_{3}W_{11}W_{31}+h^{2}_{3}W_{11}W_{13},
\end{equation}
\begin{equation}
\frac{1}{2c}\mathcal{D}W_{12} \simeq f^{2}_{3}(W_{13}W_{33}-W_{11}W_{31})+h^{2}_{3}W_{12}W_{13},
\end{equation}
\begin{equation}
\frac{1}{2c}\mathcal{D}W_{13} \simeq -f^{2}_{3}W_{13}W_{33}-h^{2}_{3}W_{13}(1-W_{33}),
\end{equation}
\begin{equation}
\frac{1}{2c}\mathcal{D}W_{21}\simeq f^{2}_{3}W_{21}W_{31}+h^{2}_{3}(W_{31}W_{33}-W_{11}W_{13}),
\end{equation}
\begin{equation}
\frac{1}{2c}\mathcal{D}W_{22} \simeq -f^{2}_{3}(W_{21}W_{31}-W_{23}W_{33})-h^{2}_{3}(W_{12}W_{13}-W_{32}W_{33}),
\end{equation}
\begin{equation}
\frac{1}{2c}\mathcal{D}W_{23} \simeq -f^{2}_{3}W_{23}W_{33}-h^{2}_{3}W_{23}(W_{33}-W_{13}),
\end{equation}
\begin{equation}
\frac{1}{2c}\mathcal{D}W_{31} \simeq -f^{2}_{3}W_{31}(1-W_{31})-h^{2}_{3}W_{31}W_{33},
\end{equation}
\begin{equation}
\frac{1}{2c}\mathcal{D}W_{32} \simeq  -f^{2}_{3}W_{32}(W_{33}-W_{31})-h^{2}_{3}W_{32}W_{33},
\end{equation}
\begin{equation}\label{DW33}
\frac{1}{2c}\mathcal{D}W_{33} \simeq f^{2}_{3}W_{33}(1-W_{33})+h^{2}_{3}W_{33}(1-W_{33}).
\end{equation}
Here, use has been made of the identities such as
$\Lambda_{11}+\Lambda_{12}=-W_{23}W_{33}$, etc. Also,
it can be verified that $\sum_{\alpha}\mathcal{D}W_{\alpha i}=
\sum_{i}\mathcal{D}W_{\alpha i}=0$.

\subsection{Case B): $f^{2}_{3} >> f^{2}_{2} \approx f^{2}_{1}$ and 
$h^{2}_{3} >> h^{2}_{2} \approx h^{2}_{1}$} 

In this case, $[F_{23},F_{31},F_{12}]\simeq (2f^{2}_{2}/\epsilon_{f})[0,0,-1]$ and
$[H_{23},H_{31},H_{12}]\simeq (2h^{2}_{2}/\epsilon_{h})[0,0,-1]$, where 
$\epsilon_{f}=f^{2}_{2}-f^{2}_{1}$ and $\epsilon_{h} =h^{2}_{2}-h^{2}_{1}$.
In addition,
\begin{equation}
[\Delta f^{2}_{23},\Delta f^{2}_{31},\Delta f^{2}_{12}]\simeq f^{2}_{3}[-1,1,0],
\end{equation}
\begin{equation}
[\Delta h^{2}_{23},\Delta h^{2}_{31},\Delta h^{2}_{12}]\simeq h^{2}_{3}[-1,1,0].
\end{equation}

The general expression for $\mathcal{D}W_{ij}$ becomes 
\begin{equation}\label{DD}
\frac{1}{2c}\mathcal{D}W_{ij}=-\eta(S^{13}_{ij}+S^{23}_{ij})+
\eta' (S_{ij}^{23}),
\end{equation}
with $S^{pq}_{ij}$ the $(p,q)$ element of $S_{ij}$,
$\eta=2f^{2}_{3}h^{2}_{2}/\epsilon_{h}$, and $\eta'=2h^{2}_{3}f^{2}_{2}/\epsilon_{f}$.
Their explicit forms are given by
\begin{equation}
\frac{1}{2c}\mathcal{D}W_{11} \simeq -(\eta \Lambda_{23}+\eta' \Lambda_{32}),
\end{equation}
\begin{equation}
\frac{1}{2c}\mathcal{D}W_{12} \simeq \eta \Lambda_{23}-\eta' \Lambda_{31},
\end{equation}
\begin{equation}
\frac{1}{2c}\mathcal{D}W_{13} \simeq -\eta' W_{13}W_{23},
\end{equation}
\begin{equation}
\frac{1}{2c}\mathcal{D}W_{21} \simeq -\eta \Lambda_{13}+\eta' \Lambda_{32},
\end{equation}
\begin{equation}
\frac{1}{2c}\mathcal{D}W_{22} \simeq \eta \Lambda_{13}+\eta' \Lambda_{31},
\end{equation}
\begin{equation}
\frac{1}{2c}\mathcal{D}W_{23} \simeq \eta' W_{13}W_{23},
\end{equation}
\begin{equation}
\frac{1}{2c}\mathcal{D}W_{31} \simeq -\eta W_{31}W_{32},
\end{equation}
\begin{equation}
\frac{1}{2c}\mathcal{D}W_{32} \simeq \eta W_{31}W_{32},
\end{equation}
\begin{equation}
\frac{1}{2c}\mathcal{D}W_{33} \simeq 0.
\end{equation}
It is seen that $\mathcal{D}(W_{13}+W_{23})\simeq 0$, $\mathcal{D}(W_{31}+W_{32})\simeq 0$,
$W_{33}\simeq $ constant, and 
$W_{11}+W_{12}+W_{21}+W_{22}\simeq $constant.

With the immediate solution for $\mathcal{D}W_{33}$,
\begin{equation}
W_{33}\approx a_{33}, 
\end{equation}
and the condition $W_{13}+W_{23}=W_{31}+W_{32}=1-a_{33}$, 
we obtain the following:
\begin{equation}
W_{13}\simeq \frac{1-a_{33}}{1-(1-\frac{1-a_{33}}{a_{13}})e^{(1-a_{33})(t-t_{0})/a_{\eta'}}},
\end{equation}
\begin{equation}
W_{23}\simeq \frac{1-a_{33}}{1-(1-\frac{1-a_{33}}{a_{23}})e^{-(1-a_{33})(t-t_{0})/a_{\eta'}}},
\end{equation}
\begin{equation}
W_{31}\simeq \frac{1-a_{33}}{1-(1-\frac{1-a_{33}}{a_{31}})e^{(1-a_{33}(t-t_{0})/a_{\eta}}},
\end{equation}
\begin{equation}
W_{32}\simeq \frac{1-a_{33}}{1-(1-\frac{1-a_{33}}{a_{32}})e^{-(1-a_{33})(t-t_{0})/a_{\eta}}},
\end{equation}
where $a_{\eta}=16\pi^{2}/(2c\eta)$ and $a_{\eta'}=16\pi^{2}/(2c\eta')$.

\subsection{Case C): $f^{2}_{3}\gg f^{2}_{2}\gg f^{2}_{1}$ and
$h^{2}_{3}\gg h^{2}_{2} \approx h^{2}_{1}$}

In this case, $[F_{23},F_{31},F_{12}]\simeq [-1,1,-1]$ and
$[H_{23},H_{31},H_{12}]\simeq (2h^{2}_{2}/\epsilon_{h})[0,0,-1]$.  In addition,
\begin{equation}
[\Delta f^{2}_{23},\Delta f^{2}_{31},\Delta f^{2}_{12}]\simeq f^{2}_{3}[-1,1,0],
\end{equation}
\begin{equation}
[\Delta h^{2}_{23},\Delta h^{2}_{31},\Delta h^{2}_{12}]\simeq h^{2}_{3}[-1,1,0].
\end{equation}
The general expression for $\mathcal{D}W_{ij}$ becomes
\begin{equation}\label{HD}
\frac{1}{2c}\mathcal{D}W_{ij}=\eta[-S_{ij}^{13}+S_{ij}^{23}]-
h^{2}_{3}[\sum^{3}_{p,q\neq 3}S_{ij}^{pq}(-1)^{p+q}].
\end{equation}
The explicit expressions are 
\begin{equation}
\frac{1}{2c}\mathcal{D}W_{11} \simeq -\eta\Lambda_{23}+h^{2}_{3}W_{11}W_{13},
\end{equation}
\begin{equation}
\frac{1}{2c}\mathcal{D}W_{12} \simeq \eta\Lambda_{23}+h^{2}_{3}W_{12}W_{13},
\end{equation}
\begin{equation}
\frac{1}{2c}\mathcal{D}W_{13} \simeq -h^{2}_{3}W_{13}(1-W_{13}),
\end{equation}
\begin{equation}
\frac{1}{2c}\mathcal{D}W_{21} \simeq -\eta\Lambda_{13}+h^{2}_{3}(W_{31}W_{33}-W_{11}W_{13}),
\end{equation}
\begin{equation}
\frac{1}{2c}\mathcal{D}W_{22} \simeq \eta\Lambda_{13}
+h^{2}_{3}(W_{32}W_{33}-W_{12}W_{13}),
\end{equation}
\begin{equation}
\frac{1}{2c}\mathcal{D}W_{23} \simeq h^{2}_{3}W_{23}(W_{13}-W_{33}),
\end{equation}
\begin{equation}
\frac{1}{2c}\mathcal{D}W_{31} \simeq -\eta W_{31}W_{32}-h^{2}_{3}W_{31}W_{33},
\end{equation}
\begin{equation}
\frac{1}{2c}\mathcal{D}W_{32} \simeq \eta W_{31}W_{32}-h^{2}_{3}W_{32}W_{33},
\end{equation}
\begin{equation}
\frac{1}{2c}\mathcal{D}W_{33} \simeq h^{2}_{3}W_{33}(1-W_{33}).
\end{equation}

The approximate solutions of $W_{33}$, $W_{13}$, and $W_{23}$ are given by
\begin{equation}
W_{33}\simeq \frac{1}{1+(a_{33}^{-1}-1)e^{-(t-t_{0})/c_{h}}},
\end{equation}
\begin{equation}
W_{13}\simeq \frac{1}{1+(a_{13}^{-1}-1)e^{(t-t_{0})/c_{h}}},
\end{equation}
\begin{equation}
W_{23} \simeq \frac{a_{23}}{[(1-a_{13})+a_{13}e^{-(t-t_{0})/c_{h}}]
[(1-a_{33})+a_{33}e^{(t-t_{0})/c_{h}}]},
\end{equation}
where $c_{h}=16\pi^{2}/(2ch^{2}_{3})$.
A special case when $\eta=2f^{2}_{3}h^{2}_{2}/\epsilon_{h}\ll h^{2}_{3}$, 
it leads to 
\begin{equation}
W_{11} \simeq \frac{a_{11}}{(1-a_{13})+a_{13}e^{-(t-t_{0})/c_{h}}},
\end{equation}
\begin{equation}
W_{12} \simeq \frac{a_{12}}{(1-a_{13})+a_{13}e^{-(t-t_{0})/c_{h}}}.
\end{equation}
\begin{equation}
W_{31} \simeq \frac{a_{31}}{(1-a_{33})+a_{33}e^{(t-t_{0})/c_{h}})},
\end{equation}
\begin{equation}
W_{32} \simeq \frac{a_{32}}{(1-a_{33})+a_{33}e^{(t-t_{0})/c_{h}})}.
\end{equation}


The RGEs and their solutions for the case of $f^{2}_{3}\gg f^{2}_{2}\approx f^{2}_{1}$ and
$h^{2}_{3}\gg h^{2}_{2} \gg h^{2}_{1}$ can be obtained from that for case C) by replacing 
$f \leftrightarrow h$.
One notes that in the literature, there exist solutions for the RGEs under different
approximate schemes, see, e.g., Refs.\cite{Balzereit:1998id,JuarezWysozka:2002kx}.

\end{document}